# DomainScope: A disease network based on protein domain connections

Alin Voskanian-kordi[1], Ashley Funai[1], Maricel G. Kann[1] *

1- Department of Biological Sciences

University of Maryland, Baltimore County, MD, U.S.A.

*-Corresponding author: mkann@umbc.edu; phone: 410-455-2258

## Abstract

Protein domains are highly conserved functional units of proteins. Because they carry functionally significant information, the majority of the coding disease variants are located on domains. Additionally, domains are specific units of the proteins that can be targeted for drug delivery purposes. Here, using information about variants sites associated with diseases, a disease network was built, based on their sharing the same domain and domain variation site. The result was 49,990 disease pairs linked by domain variant site and 533,687 disease pairs that share the same mutated domain. These pairs were compared to disease pairs made using previous methods such as gene identity and gene variant site identity, which revealed that over 8,000 of these pairs were not only missing from the gene pairings but also not found commonly together in literature. The disease network was analyzed from their disease subject categories, which when compared to the gene-based disease network revealed that the domain method results in higher number of connections across disease categories versus within a disease category. Further, a study into the drug repurposing possibilities of the disease network created using domain revealed that 16,902 of the disease pairs had a drug reported for one disease but not the other, highlighting the drug repurposing potential of this new methodology.

**Abbreviations:** UMLS- unified medical language system; MeSH- medical subject headings CUI-concept unique identified; DMDM-domain mapping of disease variation; PD- Parkinson's Disease

## 1. Introduction

Protein domains are structural units of proteins that are highly conserved and functionally identical across species[1,2]. Domains are crucial for proper protein function, as such, variations in protein domains can have consequential impact on an organism's phenotype[3,4]. Additionally, recent studies have shown that disease variations tend to cluster in the same domain sites of different genes, forming hotspots of disease variations[5-7]. The identification of these disease domain relationships provides potential for developing novel hypothesis about the molecular underpinning of diseases, that were not visible in similar gene-centric approaches. Here, we introduce a new methodology, DomainScope, that uses disease associated mutated protein domains to establish relationships between phenotypes. DomainScope provides the framework to create domain-based disease networks and thereby reveals non-obvious mechanistic and functional similarities between human disorders.

The study of relationships between human diseases is an important step towards making advancement in personalized medicine, drug repurposing, and disease research[8-10]. This relational study has been enriched by the increased availability of health and disease information. The curation of various disease relationship has culminated in the building of disease networks. These networks are built using various features of diseases, namely: gene associations and expression [11-13], protein interaction[14], pathways[15,16], etiology[17], semantic similarity[18]. As well as some hybrid approaches which integrate these features to build more robust disease networks[19,20]. These types of networks have been used to understand the origin of particular diseases[21,22] as well as drug repurposing[23-25]. Though each of these networks reveal various novel relational information, one of the remaining challenges in disease network research is introducing new molecular information, which in turn will lead to new hypotheses about molecular relationships between diseases. DomainScope does this by introducing domain disease relational information to the network medicine field.

By looking at domain-based disease connections, as we do with DomainScope, we expect to gain more insight into functional origins of phenotypes. Our method mapped variations from over 80,000 disease phenotypes from Online Mendelian Inheritance in Man (OMIM)[26], ClinVar database[27], Universal Protein Resource (UniProt)[28], and Human Gene Mutation Database (HGMD) [29] to their protein domain. Each domain was then associated to a disease Unified Medical Language System's (UMLS) Concept Unique Identifier (CUI) resulting in the identification of disease relationships among CUIs based on identities at both the domain and the domain variation site levels.

The DomainScope method yielded 533,687 disease pairs based on domain-disease associations, and 49,990 disease pairs based on disease variants associated to the same protein domain and variation site. Our domain-based results were compared with previously explored methods that built disease networks based on gene commonalities. Due to the potential of the same domain being found in variety of tissues[30], as opposed to the generally localized nature of a gene, DomainScope has resulted in identifying disease pair relationships that may have not been obvious with a gene or even pathway approach.

The pairings made by our domain approach were further explored with relation to their Medical Subject Headings (MeSH) disease categories. Contrasting with results from previous approaches many of the diseases pairs made by DomainScope stretched across disease categories as opposed to pairings within same categories. Furthermore, many of the disease pairs identified by DomainScope were missing from previous gene-based approaches. Over 8,000 of the DomainScope disease connections found by looking at identity of domain and variation site between diseases were from different genes and rarely found together in literature, indicating their non-obvious nature.

These novel connections made by using DomainScope can lead to discovery of new therapeutic methods. Of the 49,990 disease pairs 16,902 have a drug identified in one of the two disease pairs and not the other indicating their potential for drug repurposing.

## 2. Materials and Methods

### 2.1 Building the Variation dataset

The disease variant tables compiled previously in Peterson et al.[31], from OMIM, ClinVar, UniProt and HGMD were used to extract variant-disease relationships. These variants were mapped to disease/disorder concepts from UMLS[32] categorized into unique identifier concepts or CUIs. CUIs are categorically assigned to biological phenotypes and maintain the disease categorization, where applicable[15]. Additionally, MeSH categories were also incorporated into the dataset, automatically, providing disease type categorization following the methodology presented in Cirincione et al[15]. Using these two features, CUI and MeSH, allows us to categorize the dataset for a comprehensive understanding of domain disease connections.

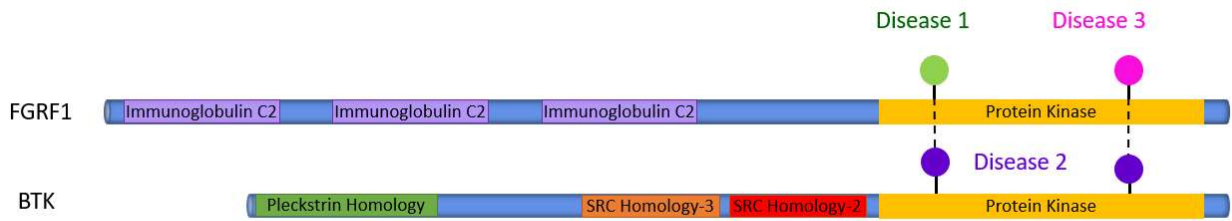

Figure 1 Two different proteins that have disease causing variations in same domain positions. This allows for a mechanism by which the diseases can be related.

### 2.2 Mapping disease variations to domains

Here, we used the approach previously used to create our Domain Mapping of Disease Mutations (DMDM) to map all human proteins, from protein data found in UniProt[33] to all protein domains from the CDD database.[34] Once each gene variation position was aligned to a representative protein (larger gene isoform from RefSeq), the HMMer's semi-global implementation[35] was used to map to a domain and position using Position Specific Mappings (PSSMs) from curated CDD domains, Pfam[36], Smart[37] and COGs[38]. Resulting in the associated domain and domain position being marked per disease variation as shown in Figure 1 on each protein.

## 2.3 Constructing the bipartite network of genes or protein domain associations to disease

The many to many relationships between gene, gene site, domain, and domain sites with disorders was created from the initial dataset of human disease variants and the mapping of these variants to a highly redundant set of protein domains (as described in previous section). As a result, the human variant dataset with domain associations contained almost 1.8 million disease gene variants entries mapped protein domains, with various information about the disease, including, the disease identifier in UMLS (CUI) with its corresponding disease name, the variant location at protein and protein domain level, and the MeSH category of the CUI.

Using the gene associations, a network of disease-disease connections was made based on gene identity and identical gene variation positions.

Similarly, two variants in the same protein domain constitute the basis of disease associations used in DomainScope at protein domain level, with a stricter requirement for the two variants to be in identical protein domain positions used in DomainScope at domain position level. This domain pairing involved 139,522 unique domain variation sites across 3,673 domains. Although some of the domains may be redundant due to the inherit redundancy between the pfam, COGs, CDD, Smart datasets. The initial database of all CUI pairs, each pairing one CUI to another based on identity of domain and domain variation contained several repetitions of CUI1-CUI2 pairing by different domain position variations. For simplicity, the connections were non-redundified to represent only one of each of the CUI-CUI pairs in the non-redundant domain and domain site disease datasets.

All disease connections were associated to their corresponding MeSH categories based on the CUI-MeSH mapping provided by UMLS.

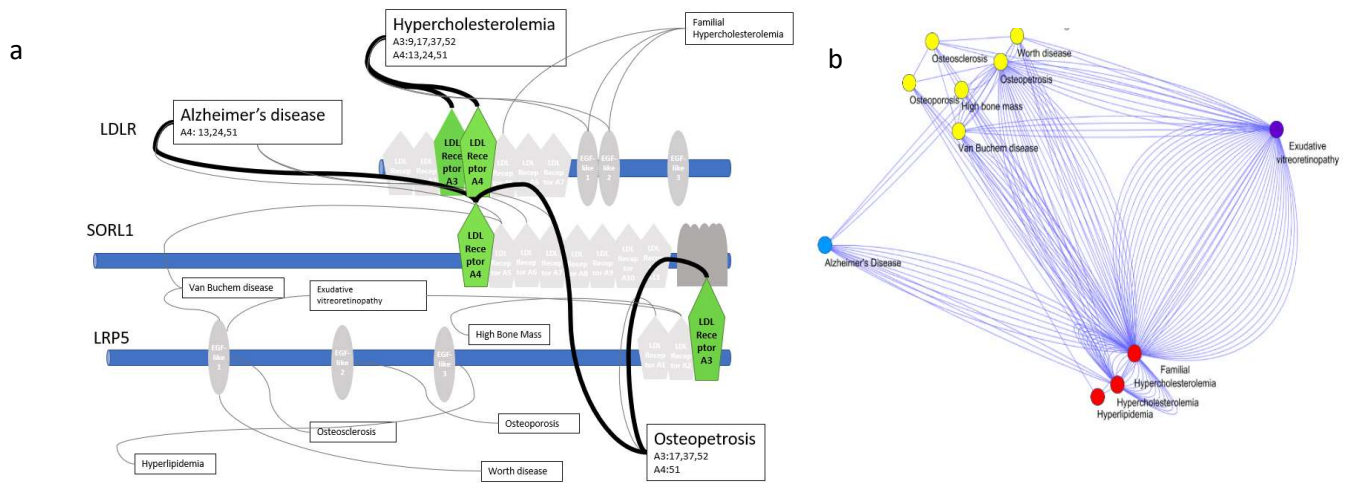

*Figure 2 Overview of network building mechanism. a- Disease connected by similar domain sites. b- Network generated by disease connections.*

## 2.4 Domain site functional analysis

To identify the functional sites that are involved in making disease pairs based on identical domain sites, we looked more closely at the functional features of the sites in the CDD domains. A subset of the CDD, the cd manually curated domains, were chosen as these are the ones with the functional site annotations from our dataset[34]. The 3.3 million disease pairs spanned 107,267 unique domain sites which were in 1,997 unique domains. To remove the effect from the high redundancy of the disease-domain mappings, i.e., most domains from a large cd cluster will map to the same disease variant, cd domains were mapped to their root cd domain and only the results of mapping functional sites to this root domain were used in our analysis. This resulted in 16,825 unique root function which were further categorized in their common type designation[39]: active, polypeptide binding site, loop, motif, nucleic acid binding, ion binding, chemical binding, postranslational modification and other (walker, box, etc).

2.5    DomainScope network visualization

Once the data was parsed and analyzed, networks were created linking the CUI-CUI pairs (Figure 2), using Cytoscape[40]. For ease of visualization, the CUI's were reduced to their MeSH categories and mapped as entities within MeSH categories (Figure 4). For this visualization, if a CUI had multiple MeSH categories it was included in all its categories. For example, Kallman syndrome 2 (C1563720) is in urological and male genital diseases (C12), female genital diseases and pregnancy complications (C13), congenital, herditary and neonatal disease and abnormalities (C16), and endocrine system diseases (C19). The thickness of the edges was calculated based on the number of connections between the two MeSH nodes, while the size of the nodes was the number of CUIs in that category. If two disease forming a CUI-CUI pair were in the same MeSH category they would be part of the self-connected loop existing on most nodes. Additionally, MeSH categories with low number of CUIs, bacterial infections and mycoses (C01), virus diseases (C02), parasite diseases (C03), disorders of environmental origin (C21) and pathological conditions, sign and symptoms (C22) were excluded from our analysis. On the other hand, congenital, hereditary and neonatal disease and abnormalities (C16), was also excluded because almost all diseases fall into this category, lacking the specificity that the other MeSH categories provide. The gene network was also visualized. Using the NetworkAnalyzer in cytoscape, the network statistics of both networks were calculated.

2.6    Heatmap of gene-based disease relationships and domain variation based

Based on the disease connections of both whole gene and domain site approaches, the combination of the corresponding MeSH categories to each disease pair, i.e., CUI1 and CUI2 was summed. Fisher's test was used to estimate each of the heatmap's cell P-values for gene and domain site CUI-CUI pairs, Figure S1. Like the disease network, if a disease was in multiple categories it was counted as a part of both categories.

2.7    PubMed disease mapping

Each CUI was used as a search term in PubMed[41] to create CUI-PMIDS database, using E-utilities tools[42]. Based on the disease pairs identified by the various approaches (gene, gene and position, domain, domain and position) the Jaccard overlap Index was estimated for each CUI pair. A criterion of minimum of 500 publications associated to each CUI and a Jacckard overlap index below 5 % was used to identify a CUI-CUI pairs of interest, i.e., non-obvious disease pairs based on publication evidence.

## 2.8 Drug- gene and disease mapping

Drug information was collected from two separate databases, DrugBank[43] and RxNorm[44]. The DrugBank database was parsed to identify drugs associated to a CUI using its gene information. With the RxNorm database, each CUI was converted to its NDF-RT identifier and the RxClass API was used to identify all drugs associated with the NDF-RT identifier using the "may_treat" relation. The resulting RxCUIs were converted to their corresponding DrugBank identifier. Based on the drugs identified through DrugBank and RxNorm, CUI-CUI pairs with common drugs were identified.

## 3. Results

Using DomainScope we were able to form disease pairs in two ways, as follows. Based on domain identity, which resulted in 533,587 disease pairs, and a more restricted way that resulted on 49,990 disease pairs associated to variants in identical protein domain positions. DomainScope results were compared with the gene-based diseaseome, which for our database resulted in 163,492 unique diseases pairs if the match was based on the diseases associations at the same gene regardless of the variant position and a total of 6,673 disease pairs matched based on identical gene positions.

The result of the analysis at the domain-site level was the identification of 49,990 disease pairs among 3,673 different domains. To understand the types of domains involved, the domain functional annotations were considered using a non-redundant set of CDD domains with functional annotation consisting of 1,997 CDD domains, though some of these 1,997 had multiple positions that related two diseases.

The results from the domain site-based analysis were further analyzed to categorize the type of functional sites where the disease connections occur. Results of such study were restricted to root domains from the CDD, which provided a non-redundant set of domains with functional annotation. Our analysis shows that a total of 16,825 functional feature sites from these CDDs were responsible for disease pairings. Figure 3 shows the frequency of these 16,825 features as categorizing by their functional groups (active, polypeptide binding site, loop, motif, nucleic acid binding, ion binding, chemical binding, postranslational modification and other (walker, box etc)). More detailed distribution of domain positions and functional annotations is available in Appendix T1.

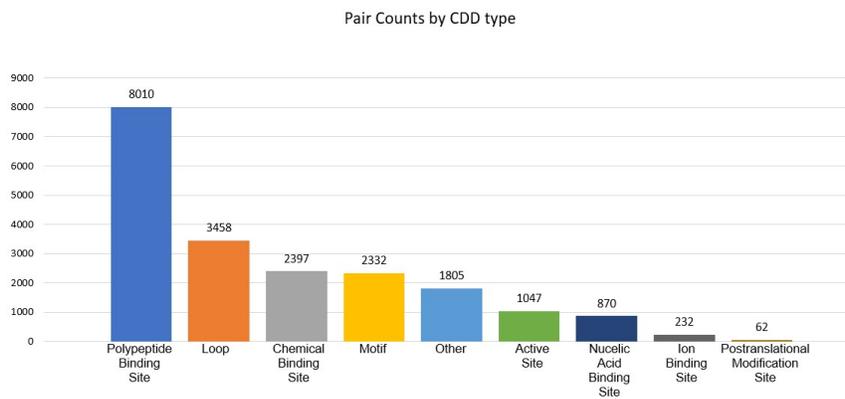

*Figure 3 Distribution of domain based on root level functional annotations*

The network of domain-variant site disease pairs, clustered by MeSH categories, is shown in Figure 4. Each node represents a MeSH category and the thickness of the lines represents number of unique CUI connected that pertained to specific MeSH categories. The size of the node indicates the number of disease or unique CUIs that pertained to that category. One CUI may belong to many MeSH categories, for Figure 4 each instance of the CUI-MeSH pair was counted individually. For a complete list of MeSH distributions see Appendix T2.

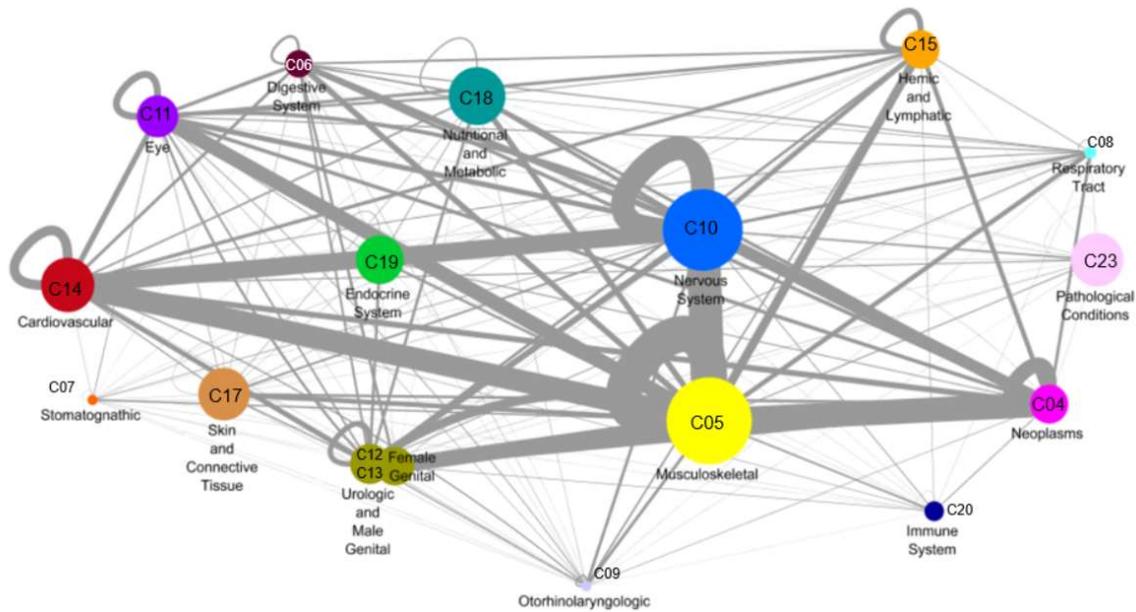

*Figure 4. Overall disease network divided into MeSH categories. Size of node represents number of diseases in that category, edge thickness represents number of diseases involved in the connections between those two diseases.*

The MeSH category connectivity of disease pairs were compared between the domain variation site approach and the gene approach. The results show, the domain variation approach has a higher tendency to connect diseases across all categories, showing a over 128,000 disease connections across disease categories, compared to 24,500 found in the gene-based network approach. The discrepancy between the totals here compared to the total connections made by gene or domains is due to two reasons, one being the lack of MeSH categorization for certain diseases and two being the multiplicity of categorization for certain diseases. Subsequently, p-values were calculated for the two approaches as shown in Figure S1. Based on the p-values, a study of the categorical connections in the gene and domain approaches shows that gene categories tend to cluster between diseases from the same categories. The gene approach revealed 20 stastically significant MeSH clusters while the domain approach has only 3 statistically significant cluster.

Furthermore, the potential drug repurposing applications for the domain-site level approach were considered. Out of the 49,990 disease pairs made at domain site-level, 11,601 shared the same drug treatment and 16,902 pairs have at least one drug treatment associated to the disease or gene for one of the disease pairs and none for the other.

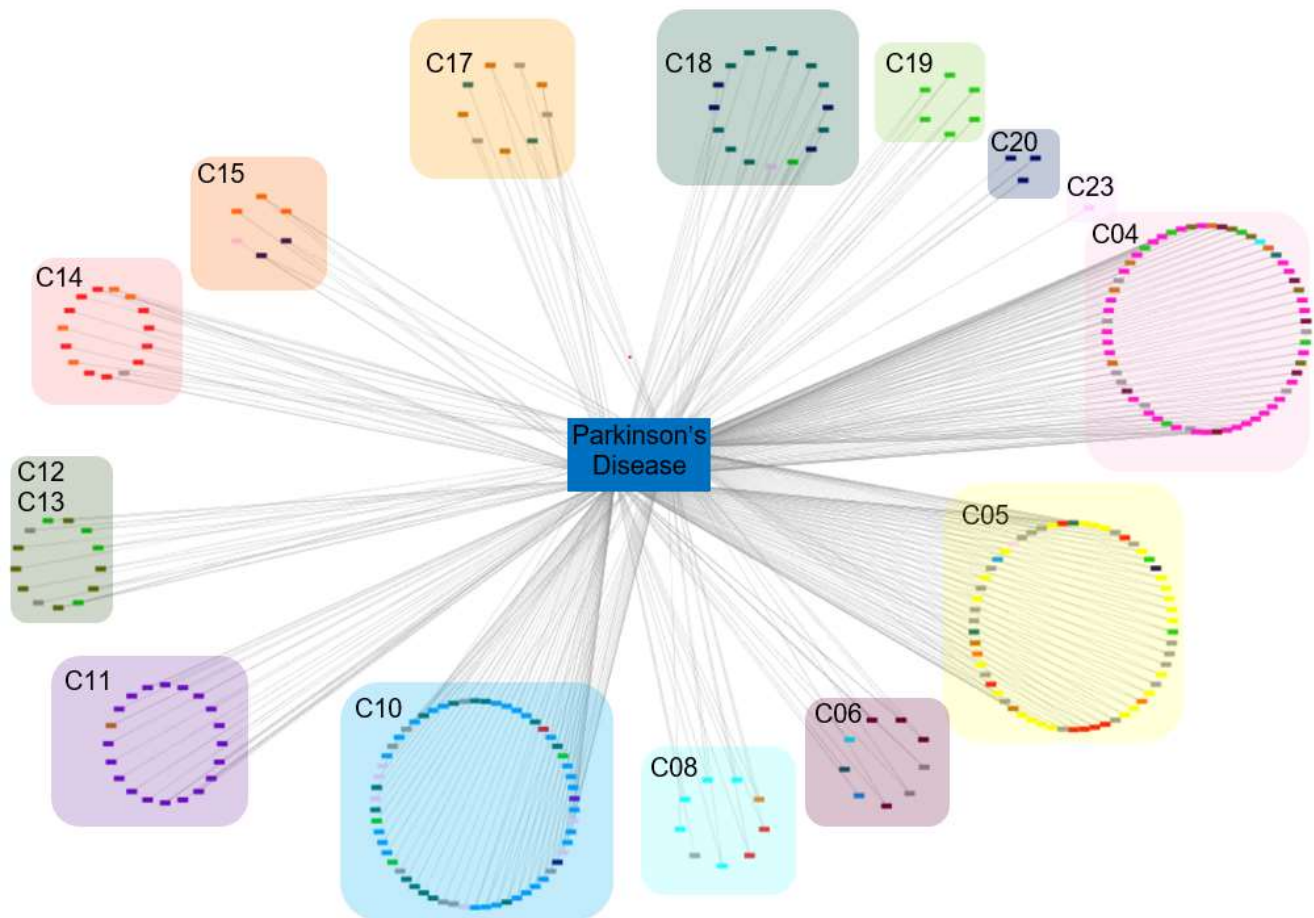

*Figure 5 Disease-Disease pairs that share one CUI that is Parkinson's Disease. The related CUI pair is organized in its MeSH category. If a CUI had two mesh categories it was colored according to the higher mesh category but placed in the group of the lower mesh category. The dark grey nodes represent CUI's that had more than two unique mesh categories.*

3.1 Case Study: Parkinson's case

To demonstrate the ability of this method in discovering non-obvious disease relationships the Parkinson's disease (PD) case was chosen, Figure 5. The center of the figure represents various CUI found in the UMLS database to represent PD. Each of the outside nodes represent a unique CUI that is connected to PD through a shared domain variation site. The CUI's are organized in their respective MeSH categories. Where a CUI has two MeSH categories it was colored according to the higher MeSH category but placed in the group of the lower MeSH category. The dark grey nodes represent CUIs that had more than two unique MeSH categories. In total there are 452 CUI pairs that resulted from this, with 202 of these pairs being novel based on the Jaccard calculations described in section 2.7. Looking at the drug reproducibility application, our analysis reveals that 300 of the disease pairs that involve PD have a known drug in the other disease pair.

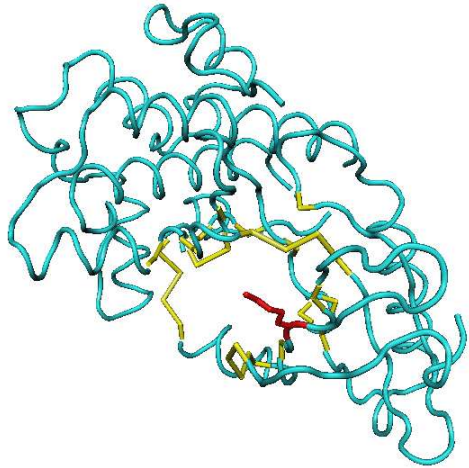

*Figure 6 The structure of the Chain A, Serine/threonine-protein Kinase Nek2 (PDB 2W5A_A) representative of the CDD cd08215 is shown with position 17 highlighted in red. Cd 08215,Position 17 was mapped to position 257 in PINK1 and 27 in TSSK2 associate*

## 4. Discussion

We introduce a novel methodology which compares the domain-variant associations of two disease phenotypes to identify disease relations. This method has great potential for discovering common molecular connections and drug targets among diseases that could lead to harvesting new therapeutical treatments and biomarkers for diagnosis and prognosis of human diseases. To better understand the value of this approach, we compare it with previously explored gene-centric approach and show the advantages of using the domain-level approach.

Using a protein domain-based approach, the number of disease-disease connections that can be made varies from 533,687, when the variant being in a common domain is used to connect diseases, to 49,990 when the more strict requirement, the variants from each diseases should be in the same domain position, is used. In contrast, when the diseases are connected using the condition that the variants are in the same gene, 163,492 disease connections are made with only 6,673 of the variants in the same gene-position.

The domain site-based disease network (Figure 4) recapitulates some of the original findings in gene-based networks, namely the high connectivity of diseases within the same broad category. However, because the network does not take into account the difference in the number of diseaseses per category, p-values were calculated which show that the gene approach has more significant p-values, most of which are in the self similarity categories. Additionally, our approach tends have a higher tendency to connect diseases across all categories, showing over 86% disease connections across disease categories, compared to 70% found in the gene-based network approach.

|  | Domain based disease network | Gene based disease network |
| --- | --- | --- |
| Number of nodes | 5287 | 5298 |
| Network heterogeneity | 1.932 | 2.224 |
| Network density | 0.004 | 0.012 |
| Connected component | 179 | 344 |
| Clustering coefficient | 0.650 | 0.792 |

*Table 1 Cytoscape NetworkAnalyzer results for the domain variant position based disease network and the gene disease network*

To compare gene-gene disease network vs the domain variation site disease network, the network metrics were calculated in cytoscape and shown in Table 1. This analysis shows that the clustering coefficient of the domain network is lower than that of the gene network. Further, the hetreogeneity of the gene network was higher, reflecting its greater hubbness.

Our results also show a larger number of "novel" disease connections found using the domain approaces versus gene based approaches. Here we defined a novel disease pair based on their co-occurrence in literature (see details in section 2.7). When using the domain-based approaches to build connections between diseases, we found 38,500 disease pairs with Jaccard index less than 5% when whole domain were used to connect and 8,618 when the more strict, same domain-site, criteria was used to build the connections. As expected, gene-based approaches, which contained less pairs across disease categories, produced a much smaller number of novel connections, 2,314 when using whole gene and only 268 when variations are in the same gene position. Table 2 highlights results for the domain-site based analysis showing the top disease pairs (22 CUI pairs) forming 100 or more novel connections.

| CUI | Disease Name | Novel Connections |
|---|---|---|
| C0035334 | Retinitis Pigmentosa | 231 |
| C0011860 | Diabetes Mellitus, Non-Insulin-Dependent | 185 |
| C0162809 | Kallmann Syndrome | 184 |
| C0339527 | Leber Congenital Amaurosis | 170 |
| C1563720 | Kallmann Syndrome 2 (disorder) | 162 |
| C0007193 | Cardiomyopathy, Dilated | 156 |
| C0001768 | Agammaglobulinemia | 141 |
| C0004352 | Autistic Disorder | 140 |
| C0019569 | Hirschsprung Disease | 140 |
| C0018553 | Hamartoma Syndrome, Multiple | 136 |
| C0010273 | Craniofacial Dysostosis | 135 |
| C0220658 | Pfeiffer Syndrome | 133 |
| C0002768 | Congenital Pain Insensitivity | 128 |
| C0221026 | X-linked agammaglobulinemia | 125 |
| C0376358 | Malignant neoplasm of prostate | 121 |
| C0030567 | Parkinson Disease | 121 |
| C0031269 | Peutz-Jeghers Syndrome | 118 |
| C0007194 | Hypertrophic Cardiomyopathy | 114 |
| C0010278 | Craniosynostosis | 114 |
| C0021655 | Insulin Resistance | 110 |
| C0020074 | HSAN Type IV | 110 |
| C0006142 | Malignant neoplasm of breast | 108 |

*Table 2 Top 22 diseases, and their corresponding CUI, that are involved in novel disease pairings using the domain variation site identity method. Novelty established using Jaccard index, based on PubMed publications*

When comparing the location of disease connections, as shown in Table 3 gene-level connections are more concentrated in Hemoglobins, while the domain connections are located in Kinase domains. Roughly 70% of disease-gene pairs are made with Hemoglobin genes, though this may be a result of the biases in the Mendelian disease set included. Furthermore, polypeptide binding sites are largely responsible for the disease pairings. These sites have great implications on potential drug docking sites[45] which could drive drug discovery.

|  | Gene | |
|---|---|---|
|  | Number of Disease Connections | Percentage of Disease Connections |
| HBB | 112101 | 0.677 |
| HBA1 | 26106 | 0.158 |
| HBA2 | 3240 | 0.020 |
| G6PD | 3003 | 0.018 |
| ALB | 1176 | 0.007 |
| SERPINA1 | 780 | 0.005 |
| HBD | 666 | 0.004 |
| TP53 | 561 | 0.003 |
| APOE | 496 | 0.003 |
| FGFR3 | 406 | 0.002 |
|  | **Domain Root** | |
| cd13968:PKc_like | 734574 | 0.898 |
| cd00267:ABC_ATPase | 28659 | 0.035 |
| cd00882:Ras_like_GT | 7154 | 0.009 |
| cd06916:NR_DBD_like | 6283 | 0.008 |
| cd01067:Globin_like | 6694 | 0.008 |
| cd00096:Ig | 6008 | 0.008 |
| cd06157:NR_LBD | 5208 | 0.007 |
| cd00172:SERPIN | 3336 | 0.004 |
| cd06534:ALDH-SF | 1487 | 0.002 |
| cd01363:Motor_domain | 1711 | 0.002 |

*Table 3 Gene, Domain Root distribution of disease pairs*

As shown in our previous discussion, one of the domain-based approaches main advantage, over gene and pathways, is its ability to connect apparently unrelated genes regardless of where they are expressed, which is also shown in the following case study. For this case study we looked specifically at PD ( Figure 5) where the connections between PD are rather well distrbuted among all disease categories. However, a majority of the disease fell in the neoplasmas (C04) and nervous system diseases (C10), forming also interesting connections with less predictable categories such as urological and male genital diaese (C12). One example of such connections was a result of a domain variation in two different genes, PINK1 and TSSK2, connecting PD and male infertility. This pairing was made through variants in the same protein domain position of the Serine-Threonin kinase (cd08215, position 17). Figure 6 highlights the structural similiarities of the variations in the genes PINK1 and TSSK2, which were made through the same domain site, regardless of the fact that the original disease data was from variants in genes without common cannonical pathways and that are expressed in different tissues, i.e., brain and testis. The structure for the catalytic domain of the Serine/Threonine Kinase ( cd08215) is represented by the Serine/Threonine-protein Kinase Nek2, chain A ( PDB code 2W5A_A), the active site, including the ATP site is highlighted in yellow with the variation in position 17 of cd08215 highlighted in red. This exemplifies the ability of our method to uncover the specific site or domain where the diseases match, providing valuable information for drug discovery. Furthermore, of all the PD connections, 300 have a discovered drug therapy, and some of these drug therapies have previously been shown to be applicable for PD. For example Zhao et al[46] showed the drug repurposing potential between PD and Asthma, Alzheimer disease, Depression, Epilepsy, Multiple scleroris, Fatigue and Immunological disorders. Of these, Multiple sclerosis, immunological disorders and eplipesy were all part of the pairings discovered by our method too, giving further justifications for the domain-based pairings resulting in drug repurposing. Other studies [47] have drawn drug repurposing connections between PD and respiratory diseases, which was also shown in our domain connections. Thus, our results shows the ability of DomainScope to identify PD pairings that are not currently reported in the literature, such as male infertility, while veryfing several of the previously discovered PD-pairs with drug repurposing value.

## 5. Conclusions

The DomainScope approach introduced in this work provides a novel framework to considerable expand the number of disease connections made based on the molecular patterns of the variants associated to the disease. In particular, the specific domain site modality provides a way to prioritize disease connections for drug repurposing or other applications for which identiying the function of the specific site aids in the molecular modelling of drugs targets and disease mechanism. The case specific study in Parkinson Disease shows that our methodology uncovers many unknown disease connections through similarities at the protein domain level addressing the need to generate new molecular hypotheses for PD, and other diseases, for which new treatements are needed.